\documentclass{IEEEtran}

\usepackage{bm}
\usepackage{psfig, epsfig}

\begin{document}

\title{Checkbochs: Use Hardware to Check Software}
\author{Sorav Bansal\footnote{This work was done with guidance and
support from Mendel Rosenblum, Professor of Computer Science, Stanford
University}}

\twocolumn[\begin{@twocolumnfalse}


\maketitle
\thispagestyle{empty}
\begin{abstract}
In this paper, we present a system called Checkbochs, a machine simulator
that checks rules about its guest operating system and
applications at the hardware level. The properties to be checked can
be implemented as `plugins'
in the Checkbochs simulator. Some of the properties that were checked
using Checkbochs include null-pointer checks, format-string vulnerabilities,
user/kernel pointer checks, and race-conditions. On implementing these
checks, we were able to uncover previously-unknown bugs in widely used
Linux distributions. We also tested our
tools on undergraduate coursework, and found numerous bugs.
\end{abstract}
\begin{keywords}
Software Reliability, Machine Simulation, Type Systems
\end{keywords}

\end{@twocolumnfalse}]

\section{Introduction}
The use of software in many mission critical applications has led computer
scientists to focus on software reliability, more than ever before.
Many ways have been proposed to make software more reliable, secure
and bug-free. Broadly, these techniques can be grouped into static
compile-time techniques and dynamic run-time techniques.

Static compile-time techniques analyze program source code to detect
errors. This method,
though very effective cannot take advantage of the information available
at run-time. Dynamic runtime techniques on the other hand mostly
rely on an interpreted environment, or instrumentation at source-code
or binary level. For this reason, dynamic runtime techniques have seldom been
used to test the correctness of an operating system. Instrumenting the
source code or binary of an operating system requires clever engineering
effort and deep understanding of the OS internals. Further, binary
instrumentation requires thorough understanding of the instruction set
architecture. And even when the instrumentation is complete and correct,
the execution of instrumented code can change the behavior of the OS
indeterministically due to a different interleaving of concurrent threads.

In this paper, we present a dynamic runtime checker called Checkbochs which
has been implemented using an existing x86 machine simulator called Bochs
\cite{bochs}. Since each machine instruction in a simulator is interpreted,
it allows us to do complex runtime analyses without the pitfalls associated
with instrumentation. By tracking data flow at runtime, and checking for
some known properties, we found interesting bugs in two widely used
Linux distributions. We also implemented a dynamic data race detector
inside the simulator to be able to find data races in operating systems. As
a case study, we checked an academic operating system called Pintos which
was built as part of undergraduate coursework at Stanford.

Some of our contributions are:
\begin{itemize}
\item Checkbochs allows dynamic runtime analysis by implementing rules
to be executed while interpreting instructions in the guest software.
\item Checkbochs can identify bugs across all layers of software running
on the system. It does not require prior knowledge of the applications or
their source code before reporting violations.
\item Checkbochs does not change application or OS behavior in any way since
all its operations occur at the hardware level. In fact, Checkbochs can be
used on an existing disk image.
\item Implementing new runtime rules in Checkbochs is easy and straightforward
\end{itemize}
This paper is organized as follows. Section \ref{sec:dataflow} explains
the Checkbochs dataflow analysis framework that can be extended to check
various system rules at runtime. Section
\ref{sec:eraser} describes the implementation of one such rule -
the dynamic data race detector. Section \ref{sec:relwork} discusses
the related work, and finally Section \ref{sec:conclusion} concludes.

\section{Dynamic Data Flow Analysis}
\label{sec:dataflow}
In our dynamic data flow analyses, we wish to infer types
of machine values and then check for violations of type-properties
at runtime. For example, if two file-descriptors are multiplied at
any point in the execution of the system, it signals an error.

Checbochs provides a framework, which allows users to specify
data types and corresponding system rules to be checked during
runtime. In this section, we describe the implementation of
this framework and also the results obtained by implementing
some rules.

\subsection{Implementation}
To track types of machine values, Checkbochs maintains a {\em shadow
machine state}
The shadow state consists of the shadow registers, instructions
and shadow physical memory. Since the instructions use virtual memory addresses,
we also implemented a virtual-memory translation logic in the shadow space.

Data flow is monitored by associating all data transfer and arithmetic
instructions ({\tt load}, {\tt store}, {\tt move}, {\tt add}, etc.) with
their counterparts in shadow space. For example, a {\tt move dest, src}
instruction
would also cause the contents of {\tt src} to be copied to {\tt dest} in
shadow space. Using this data flow framework, we tag data values with
their {\em types} in shadow space and analyse their flow across the system.
Some of the types that we inferred using Checkbochs are tabulated in
Table \ref{tab:types} with their method of inference.

\begin{table}[htb]
\begin{center}
\begin{tabular}{|p{3.5cm}|p{4.5cm}|}
\hline
Method Of Inference & Types \\
\hline
\hline
Linux System Call Interface & User Pointers, File Descriptors, Strings \\
\hline
x86 Instruction Operands & I/O address, I/O bytes, NullChecked, Memory Address, Code/Data\\
\hline
Planted Shadow Instructions in Guest Software & malloc/fopen return value,
user pointer safety\\
\hline
\end{tabular}
\label{tab:types}
\caption{\small{Type values and their Method of Inference at the Hardware Level}}
\end{center}
\end{table}

Sometimes, the flow of type information is not directly associated with
the corresponding data flow. For example, consider the sample code of
the {\tt read()} system call in Figure \ref{fig:sysread}.
\begin{figure}[htb]
\begin{minipage}[]{\columnwidth}
{
\begin{small}
\begin{verbatim}
  ssize_t sys_read(...,char *buf,...) {
      char *tmp = buf;
      if (!access_ok(tmp)) fail();
      ...
      *buf = 42; /* safe? */
  }
\end{verbatim}
\end{small}
}
\end{minipage}
\caption{\label{fig:sysread}
The system call parameter {\tt buf} is an unsafe user pointer. However, since
a derived variable {\tt tmp} has been checked for safety, the assignment
to {\tt *buf} should be considered legal.
}
\end{figure}
At the entry of the function, {\tt buf} is an {\em unchecked user pointer}. The
assignment statement {\tt tmp=buf} changes the type of {\tt tmp} to
an unsafe user pointer too. Subsequently, {\tt tmp} goes through the
appropriate check and its type is changed to a safe user pointer. However,
the result of this check is not reflected on the type of {\tt buf} and the
line {\tt *buf=42} will incorrectly flag a warning.

\begin{figure}[htb]
\centerline{\epsfig{figure=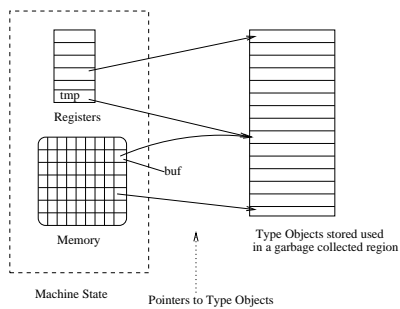,width=0.6\columnwidth}}
\caption{\label{fig:indirection}{\bf Pointers to Type Objects}.
Both {\tt tmp} and {\tt buf} point to the same type object, after
assignment of {\tt tmp} to {\tt buf}.
Hence, any change to the type of {\tt tmp}
will also reflect in the type of {\tt buf}.
}
\end{figure}

We solved this problem by using an extra level of indirection in the shadow
space. Instead of holding the type value, the shadow state holds pointers
to type objects. Hence, an assignment statement like {\tt tmp=buf} causes
the pointer to {\tt buf}'s type object to be copied to {\tt tmp}'s shadow
state. Since both {\tt buf} and {\tt tmp} now point to the same type
object, any change in {\tt tmp}'s type is reflected in {\tt buf}'s type
and vice versa. This extra level of indirection is also illustrated in
Figure \ref{fig:indirection}

\subsection{Results}
We ran two different disk images on Checkbochs: {\tt debian-3.0r1} and
{\tt gentoo-2004.3}. Both these disk images were obtained after freshly
installing the latest available distribution CDs by these vendors. Using
our dataflow framework, we check for flavors of copyin/copyout bugs, improper
handling of {\tt malloc()}, {\tt fopen()} functions and formatstring
vulnerabilities. The results obtained by performing these checks are
described in detail in the following sections.\\
\subsubsection{CopyIn/CopyOut Bugs}
A user value passed as a system call argument must be checked through one
of the copyin/copyout functions before getting dereferenced. Failure to do
so, opens a port of attack, whereby an attacker can crash the kernel,
or worse be able to write his own data at a specific kernel address. Another
variation of this rule is that a user pointer should never be dereferenced
in the kernel with disabled interrupts.

Implementing this rule using our data flow framework was straightforward. All
system call parameters are tagged as {\em unchecked user values}. Any check
on the user value through one of the copyin/copyout functions cause the
type of the user value to change from {\em unchecked} to {\em checked user
value}. A warning is
flagged if an unchecked user value is dereferenced in the kernel or
a user value is dereferenced with disabled interrupts.
 
Using these rules, we identified one copyin/copyout bug in the {\tt poll}
system call of the linux kernel. In this instance, the user pointer was
checked for read access, while a write operation was performed on it. This
can be a security flaw on many architectures, including Intel 386.
The submitted bug was acknowledged by Alan Cox on the Linux Kernel
Mailing List \cite{lkml}.\\

\begin{figure*}[t]
\begin{center}
\begin{tabular}{|c|c|c|c|}
\hline
Rule & Kernel Bugs & Application Bugs & False Positives \\
\hline
\hline
Pointers/File Descriptors checked against NULL & 6 & 16 & 0 \\
\hline
CopyIn/CopyOut Bugs & 1 & - & 4 \\
\hline
Untrusted IO in FormatString & 0 & 0 & 1 \\
\hline
\end{tabular}
\caption{\label{fig:results}
{\bf Summary}: The table summarizes the results obtained on implementing
certain rules
and checking two widely used Linux distributions. Besides these rules,
we implemented the race-detection algorithm in Checkbochs which uncovered
numerous bugs in undergraduate coursework.
}
\end{center}
\end{figure*}

\subsubsection{Improper handling of {\tt malloc()}, {\tt fopen()} return values}
Any use of {\tt glibc} functions like {\tt malloc()} and {\tt fopen()} must be
accompanied by a null check before they are used. To check this rule,
we tagged the return values of these functions until they were checked against
{\tt null}. Implementing this rule, we found many instances of violation
in the kernel and common applications like {\tt ps}, {\tt grep}, {\tt fsck},
and {\tt swapon}. The violations in the kernel were found in the
IDE device driver of {\tt linux-2.4.18}. In all, after minimal testing,
we found 16 bugs in user-level
software and 6 bugs in the kernel. All these bugs were accepted, and
subsequently fixed.\\

\subsubsection{Formatstring Vulnerabilities}
A formatstring vulnerability  \cite{formatstring}
is caused due to a design misfeature in the C
standard library combined with problematic implementation of variable
argument functions. A value from an untrusted source (such as the network)
should not be used without proper checks inside the formatstring argument
of the {\tt printf} family of functions. Failure to do so, can lead to a
complete compromise of security, when combined with other bugs. To test
applications, we mounted an NFS partition on the guest system and ran a
number of applications on it. Although, we hoped to obtain some bugs in
applications running on data in the untrusted NFS partition, we obtained
only one false positive in the {\tt sendmail} program. In this case,
each character of the string received over the network was carefully
checked before using it in the formatstring. Since our rule does not
implement a way of finding out whether a value has been appropriately
checked for safety, it reports a false warning.\\\\

In essence, we believe that machine simulation allows us to track
data flow across many different applications, which would have been
difficult (if not impossible) to do with instrumentation techniques. We
are investigating other uses of this dataflow framework in determining
interesting system properties. Recently, dynamic dataflow tracing was
used to determine lifetime of sensitive data like passwords \cite{taintbochs}.
The work revealed the startling result that sensitive information could
live in the system for a long time in places like system caches and device
buffers (places which are not in the control of the application vendor).

\section{Race Conditions}
\label{sec:eraser}
Multi-threaded programming is difficult and error-prone. It is easy to make
a mistake that produces a data race, yet it can be extremely hard to
locate this mistake during debugging. A very effective dynamic data
race detection algorithm is the lockset algorithm, first proposed in
a tool called Eraser \cite{eraser}. Eraser used binary rewriting techniques
to monitor every shared memory reference and verify that consistent locking
behavior is observed. The lockset algorithm enforces the simple locking
discipline that every shared variable is protected by some lock. Since
there is no way of knowing which locks are intended to protect which
variables, this protection relation can be inferred from the execution
history. Figure \ref{fig:lockset} summarizes the
lockset algorithm. More details on the algorithm can be found
at \cite{eraser}.

We implemented the lockset algorithm in Checkbochs.
On testing undergraduate coursework \cite{cs140}, we found numerous
bugs in assignments that had received a near-perfect score. We also
found one benign race in the base operating system {\tt pintos} \cite{pintos}
provided to the students.
\begin{figure}[htb]
\begin{minipage}[]{\columnwidth}
{
\begin{small}
\hspace{10mm}
Let $locks\_held(t)$ be the set of locks held by thread $t$

\hspace{10mm}
For each $v$, initialize $lockset(v)$ to the set of all locks

\hspace{10mm}
On each access to $v$ by thread $t$,

\hspace{10mm}
\ \ set $lockset(v)$ := $lockset(v) \bigcap locks-held(t)$;

\hspace{10mm}
\ \ if $C(v)$ = $\{\}$, then issue a warning.
\end{small}
}
\end{minipage}
\caption{\label{fig:lockset}
  {\bf The Lockset Algorithm}
}
\end{figure}

\section{Related Work}
\label{sec:relwork}
There has been a growing impetus on software reliability and security in
recent years. Researchers have considered many ways to perform 
post-production checks
in software.

Static compile-time analysis with programmer written compiler-extensions
was used to catch around 500 bugs in the linux kernel
\cite{metacompilation1}, \cite{metacompilation2}.
Using static data flow analysis and domain specific knowledge, many bugs
were found in the heavily audited kernel. Ways have also been suggested
to automatically detect anomalies as deviant behavior in the source
code \cite{deviant}. Most of the bugs checked by static analysis are
local to a single file, sometimes even local to a single procedure. This is
due to the complexity involved in performing global compile time
analysis. This limits the power of static analysis tools to {\em surface
bugs}. Our approach, on the other hand, can track data flow across many
different software components possibly written by different vendors and
can thus target a different variety of errors. However, static analysis
has the huge advantage of being able to check all possible code paths, while
our execution-driven approach can only check bugs along the path of
execution in the system.

Recently, model checking was used to find serious file system errors
\cite{modelcheckingOSDI04}. Using
an abstract model and intelligent reduction of the state space, they could
check for errors which would have required an exponential number of
search paths through traditional testing. Model checking can check for
deeper semantic bugs than possible with static compile-time analysis. We
intend to use similar ideas to model check entire system images, thus allowing
us to search a larger number of execution paths while performing our
shadow machine analysis. One of the obstacles in this direction is the
slow speed of machine simulation that makes execution of speculative
paths almost infeasible.

Shadow machine simulation has been previously used to perform taint analysis
to determine
the data lifetime of sensitive data \cite{taintbochs}. This work reported a
startling observation that sensitive data like passwords and credit card
numbers may reside in computer's memory and disk long after the user
has logged out. Such leaks occur at caches, I/O buffers, kernel queues,
and other places which are not under the control of the application
developer. Our work uses a similar taint analysis by marking all bytes received
over the network as untrusted and checking if they are used in unwanted
ways (eg. formatstring).

Recently, \cite{valgrindtaint} used taint-analysis on untrusted data to
check for security violations such as buffer overflows and formatstring
attacks in applications. By implementing a valgrind
skin, they were able to restrict the overhead of their taint-analysis
tool to 10-25x. Considering that the computation power is relatively cheap,
they suggest using their tool in production runs of the software. This will
detect and prevent any online attacks on the system.

\section{Conclusion}
\label{sec:conclusion}
We present a novel technique to finding bugs and security holes in system
software. Our technique can check for bugs across all layers
of software, from the OS to the application.
Our approach has very low false positive and false
negative rates. This technique can be especially
very useful in expediting the process of discovering bugs during software
testing.

We conjecture that shadow machine simulation, combined with speculative
execution (such
as model-checking) can yield a huge number of bugs. While the slow speed
of machine simulation is an impediment to this approach, we are considering
using virtual machine environments to achieve the same objective.

\section{Acknowledgements}
This work was done under the able guidance
and support from Mendel Rosenblum, Professor of Computer Science,
Stanford University. The author would also like to thank Ben Pfaff
for very insightful discussions, help with testing Checkbochs on
{\tt pintos}, and making Checkbochs available to CS140 students.

\bibliographystyle{plain}
\bibliography{bibliography}

\begin{thebibliography}{66.}
\small{
\bibitem{metacompilation1} D. Engler, B. Chelf, A. Chou, S. Hallem. Checking
System Rules Using System-Specific, Programmer Written Compiler Extensions,
Appeared in OSDI 2000
\bibitem{metacompilation2} K. Ashcraft, D. Engler. Using Programmer-Written
Compiler Extensions to Catch Security Holes,
IEEE Security and Privacy 02
\bibitem{deviant} D. Engler, D. Chen, S. Hallem, A. Chou, B. Chelf. Bugs
as Deviant Behavior: A General Approach to Inferring Errors in System
Code. Appeared in SOSP 01
\bibitem{modelcheckingOSDI04} J. Yang, P. Twohey, D. Engler, M. Musuvathi.
Using Model Checking to Find Serious File System Errors. Appeared in OSDI 04
\bibitem{bochs} Bochs: The cross platform IA-32 emulator.
{\tt http://bochs.sourceforge.net/}.
\bibitem{taintbochs} J. Chow, B. Pfaff, K. Christopher, M. Rosenblum.
Understanding Data Lifetime via Whole-System Simulation, USENIX Security
Symposium 2004.
\bibitem{userKernel} Rob Johnson and David Wagner. ``Finding User/Kernel
Pointer bugs with Type Inference,'' 13th USENIX Security Symposium 04
\bibitem{formatstring} Tim Newsham. ``Format String Attacks,'' Guardent,
Inc. September 2000. {\tt http://www.guardent.com/docs/FormatString.PDF}
\bibitem{lkml} Linux Kernel Mailing List. {\tt http://lkml.org}
\bibitem{crash} Mission Critical Software's {\tt crash}. http://www.missioncritical.com/support/crash.php
\bibitem{gdb} GNU Debugger. http://www.gnu.org/software/gdb/gdb.html
\bibitem{cs140} Undergraduate Course CS140. http://cs140.stanford.edu
\bibitem{pintos} Pintos: instructional operating system.
http://www.stanford.edu/class/cs140/projects/
\bibitem{bugs} A list of the bugs reported and fixed using Checkbochs. http://www.stanford.edu/people/sbansal/bugs.html
\bibitem{valgrindtaint} J. Newsome and D. Song. Dynamic taint analysis for
automatic detection, analysis and signature generation of exploits on
commodity software, to appear in Network and Distributed System Security
Symposium 05.
\bibitem{eraser} S. Savage, M. Burrows, G. Nelson, P. Solbovarro and T.
Anderson. Eraser: A dynamic data race detector for multi-threaded programs. In
Proceedings of the 19 Sixteenth Symposium on Operating Systems Principles, 1997
}
\end{thebibliography}

\end{document}